\documentclass[twocolumn,prb,aps]{revtex4}
\usepackage{graphicx}
\renewcommand{\vec}[1]{{\mathbf{#1}}}
\newcommand{\beq}{\begin{eqnarray}}
\newcommand{\eeq}{\end{eqnarray}}
\begin{document}
\draft

\title
{Theory of the Luttinger Surface in Doped Mott Insulators}

\author{Tudor D. Stanescu}
\affiliation{Condensed Matter Theory Center, Department of Physics, University of Maryland, College Park, Maryland 20742-4111}
\author{Philip Phillips and Ting-Pong Choy}
\affiliation{Loomis Laboratory of Physics,
University of Illinois at Urbana-Champaign,
1100 W.Green St., Urbana, IL., 61801-3080}

\begin{abstract}
We prove that the Mott insulating state is characterized by
a divergence of the electron self energy at well-defined values of momenta in the first Brillouin zone.  When particle-hole symmetry is present, the divergence obtains at the momenta of the Fermi surface for the corresponding non-interacting system.   Such a divergence gives rise to a surface of zeros (the Luttinger surface) of the single-particle Green function and offers a single unifying principle of Mottness from which pseudogap phenomena, spectral weight transfer, and broad spectral features emerge in doped Mott insulators. We also show that only when particle-hole symmetry is present does the volume of the zero surface equal the particle density.  We identify that the general breakdown of Luttinger's theorem
in a Mott insulator arises from the breakdown of a perturbative expansion for the self energy in the single-particle Green function around the non-interacting limit. A modified version of Luttinger's theorem is derived for special cases.
\end{abstract}
\maketitle

\section{Introduction}
In the absence of disorder, electronic states in insulators fail to carry current either because the band is full (a band insulator) or a gap opens at the chemical potential. The latter is indicative
of either a structural transition in which a partially filled band splits by doubling the unit cell or a Mott state in which strong on-site correlations create a charge gap in a half-filled band, as in transition metal oxides, the high-temperature copper-oxide superconductors a case in point.  While the origin of the Mott gap is clear, the mechanism for
the bifurcation of the half-filled band into upper and lower Hubbard bands remains controversial\cite{laughlin}.  We show here that the Mott gap leads to a divergence of the self-energy in momentum space, which in the case of particle-hole symmetry
lies on the {\bf non-interacting Fermi surface}.  Such a divergence prevents the renormalized energy band from crossing the Fermi energy and hence is ultimately
the mechanism by which Mott insulators insulate.   Further, this divergence leads to a vanishing of the single-particle Green function thereby defining a surface of zeros in
 the first  Brillouin zone.  We demonstrate that the zeros account for numerous anomalous properties of the normal state of the cuprate superconductors. The zeros found here obtain fundamentally from
strong coupling Mott physics\cite{stanescu4} and should be distinguished from those arising from
weak-coupling symmetry-breaking instabilities\cite{rice} of the Fermi surface.

That the Fermi surface of the non-interacting system is in some way connected
with strong-coupling Mott physics (that is, Mottness) has not been anticipated.  For Fermi liquids, such a correspondence is natural.
 Fermi liquid theory\cite{landau} rests on the seemingly simple conjecture that
the number of bare electrons at a given chemical potential equals the number of Fermi excitations (quasiparticles) in the interacting system regardless of the strength of the interactions.  Quasi-particle excitations
are identified by simple poles of the single-particle, time-ordered Green function, $G(\vec p,\epsilon)$.  Hence,
for a Fermi liquid, the Landau conjecture, proven by Luttinger\cite{luttinger}, implies the existence of a surface (the Fermi surface) in momentum space
at which $\Re G(\vec p, \epsilon=\epsilon_F)$ changes sign by passing through infinity.
 In systems lacking quasiparticles (no poles), such as insulators, the Landau correspondence between the particle density and quasiparticle excitations
breaks down. In fact, as uncloaked recently\cite{rice,dzy,essler}, Luttinger's theorem is not necessarily invalidated
 when quasiparticles are absent.  The suggestion\cite{dzy,essler,volovik} is
that the particle density
\beq\label{density}
\frac{N}{V}=2\int_{G(p,0)>0}\frac{d^3p}{(2\pi)^3}\equiv n_{\rm Lutt}
\eeq
is properly defined through an integral in momentum space over a region
where the single particle propagator
is positive. An explicit assumption in Eq. (\ref{density}) is that the imaginary part of the self-energy vanishes at the chemical potential\cite{schmalian}.  Sign changes occur at poles or at zeros of the Green function.
  Should Eq. (\ref{density}) hold, then the volume enclosed by the surfaces of zeros and poles yields the particle density.

In this paper, we offer a criterion for the location of the surface of zeros
and show explicitly that the volume enclosed by the zeros yields the electron density only when a Mott insulator has particle-hole symmetry. Our proof of the latter makes general the perturbative arguments made recently for a Kondo insulator\cite{rosch}.  In the absence of particle-hole symmetry, there is an additional contribution to the electron density in
Eq. (\ref{density}) which arises from the breakdown of perturbation theory. We show explicitly that the breakdown of Luttinger's theorem does not obtain from
a $T=0$ regularization of divergent integrals as has been claimed\cite{dzy,rosch} but rather arises anytime the self energy cannot be obtained perturbatively around the non-interacting limit.  Our results are in agreement with the mechanism proposed by Altshuler, et. al.\cite{altshuler} in the context of the breakdown of Luttinger's theorem in the presence of a spin-density wave.
Finally, we demonstrate that models which project out the high-energy scale at half-filling lose the surface of zeros.  As a consequence, the quasiparticle weight need not\cite{noncomm} be the same in the Hubbard and projected schemes such as the t-J model. The evolution of the surface of zeros in the pseudogap phase is also delineated.

\section{Zeros in a Mott Insulator}
As our starting point, we consider the simplest model which captures the physics of Mott insulators,  the Hubbard model,
\beq
H = -\sum_{i,j,\sigma} t_{ij}c_{i\sigma}^{\dagger}c_{j\sigma} +
U\sum_{i} n_{i\uparrow}n_{i\downarrow}-\mu\sum_{i\sigma} n_{i\sigma},
\eeq
in which electrons hopping on a lattice between neighbouring sites with amplitude $t_{ij} = t\alpha_{ij}$ and chemical potential $\mu$ pay an energy cost $U$ anytime they doubly occupy the same site.  The operator $c_{i\sigma}$ ($c_{i\sigma}^\dagger$) annihilates (creates) an electron on site $i$ with spin $\sigma$ and $\mu$ sets the chemical potential.  The quantity of interest
is the single-particle retarded Green function,
$G^{\rm ret}_\sigma(i,j,t,t')=-i\theta(t-t')\langle\{c_{i\sigma}(t),c_{j\sigma}^\dagger(t')\}\rangle$,
in particular, its momentum and energy Fourier transform, $G^{\rm ret}_\sigma(k,\epsilon)=FT G^{\rm ret}_\sigma(i,j,t,t')$, where $\{a,b\}$ indicates the anticommutator of $a$ and $b$ and $\theta(x)$ the Heaviside step function which is non-zero only if its argument is positive. The quantity that is directly observable experimentally through ARPES is the spectral function,
$A_\sigma(\vec k,\epsilon)=-\Im G^{\rm ret}_\sigma(\vec k,\epsilon)/\pi$.
Summed over momentum, the spectral function defines the single-particle density of states.   The causal nature of the Green function permits it to be constructed entirely from its imaginary part
\beq
G^{\rm ret}_\sigma(\vec k,\omega)=\int_{-\infty}^\infty d\omega' \frac{ A_\sigma(\vec k,\epsilon')}{\epsilon-\epsilon'+i\eta}
\eeq
through the standard Hilbert representation.  For a Mott insulator, a gap of order $U$ occurs in the spectral function.  We will take the gap to have a width $2\Delta$ centered about $0$.  As we consider the general case in which symmetry-breaking plays no role in the gap, simply strong electron-correlations, our conclusions regarding the zeros are applicable to the mechanism proposed by Mott.  Within the gap, $A(\vec k,\epsilon)=0$.
This is a necessary condition for any gap.  Consequently, in the presence of a gap, the real part of the Green function evaluated at the Fermi energy
reduces to
\beq\label{real}
R_\sigma(\vec k,0)=-\int_{-\infty}^{-\Delta_-}d\epsilon'\frac{ A_\sigma(\vec k,\epsilon')}{\epsilon'}-\int_{\Delta_+}^\infty
d\epsilon' \frac{A_\sigma(k,\epsilon')}{\epsilon'}
\eeq
as pointed out by Dzyaloshinskii\cite{dzy}.  At half-filling, Mott insulators have half the spectral weight above the gap.  Hence, it is, possible,
in principle, that the real part of the Green function vanishes along some momentum surface.  However, no criterion has been given for the location of such a surface.  Indeed, the nature of the Mott transition has been studied extensively in
$d=\infty$ using dynamical mean-field theory (DMFT)\cite{dmft}.  In DMFT ($d=\infty$),
the self-energy diverges at $\omega=0$  for all momenta.  Hence, in $d=\infty$, there is no Luttinger surface.  For finite dimensional systems, it is not known what becomes of the unphysical momentum-independent $1/\omega$ divergence of the self energy in
$d=\infty$   The new feature that this work brings into focus is the divergence of the self energy along a continuously connected momentum surface in the first Brillouin zone as the defining
 feature of a Mott insulator in finite dimensions.

\subsection{Particle-Hole Symmetry}
We now prove that when particle-hole symmetry is present, the
spectral function is an even function of frequency at the
non-interacting Fermi surface. As a consequence, Eq. (\ref{real}) is
identically zero along that momentum surface.  To proceed, we
consider a general particle-hole transformation,
\beq\label{ph}
c_{i\sigma}\rightarrow e^{i\vec Q\cdot \vec r_i}c^\dagger_{i\sigma},
\eeq
of the electron annihilation operator. That the Hamiltonian
remain invariant under this transformation places constraints on
both $\vec Q$ and the chemical potential. The Hubbard model with
nearest-neighbour hopping remains invariant under Eq. (\ref{ph}) for
$\vec Q=(\pi,\pi)$ and $\mu=U/2$ .  The latter is the value of the
chemical potential at half-filling, the Mott state.  Transforming
the operators in the Green function according to Eq. (\ref{ph}) and
keeping the chemical potential fixed at $\mu=U/2$ leads to the
identity
\beq\label{aspec}
A_\sigma(\vec k,\omega)=A_\sigma(-\vec
k-\vec Q+2n\pi,-\omega).
\eeq
Hence, the spectral function is an even function of frequency for $\vec k=\vec Q/2+n\pi$. Consider one
dimension and nearest-neighbour hopping.  In this case, the symmetry
points are $\pm\pi/2$, the Fermi points for the half-filled
non-interacting band.  In two dimensions, this proof is sufficient
to establish the existence of only two points, not a surface of
zeros.  To determine the surface, we take advantage of an added
symmetry in higher dimensions.  For example, in two dimensions, we
can interchange the canonical x and y axes leaving the Hamiltonian
unchanged only if the hopping is isotropic.  This invariance allows
us to interchange $k_x$ and $k_y$ on the left-hand side of Eq.
(\ref{aspec}) resulting in the conditions
\beq\label{cond1}
k_y=-k_x-q+2n\pi
\eeq
and by reflection symmetry
\beq\label{cond2}
-k_y=-k_x-q+2n\pi,
\eeq
where $\vec Q=(q,q)$.   For nearest neighbour hopping, the resultant
condition, $k_x\pm k_y=-\pi+2n\pi$, is the solution to $\cos
k_x+\cos k_y=0$, which defines the Fermi surface for the
non-interacting system. If only next-nearest neighbour hopping is
present, the value of the wave vector that leaves the kinetic energy
term unchanged after a particle-hole transformation is $\vec
Q=(\pi,0))$ or $(0,\pi)$. Coupled with Eq. (\ref{aspec}) and
reflection symmetry we also obtain the Fermi surface of the
non-interacting system.  The interactions need not be the
local on-site repulsion in the Hubbard model for the surface of
zeros to persist.  Nearest-neighbour interactions of strength $V$
depend only on the particle density and hence are independent of $Q$
under a particle-hole transformation. Such interactions renormalise
the chemical potential from $U/2$ (on-site interactions) to
$(U+2V)/2$ at the particle-hole symmetric point.

However, an implicit assumption in our proof which allows for the interchange of the momenta in Eq. (\ref{aspec}) is that the hopping is isotropic.
Nonetheless, the result we have obtained is independent of the isotropy of the hopping.  That is, our proof applies equally when the band structure is of the form $t(k)=t_x\cos k_x +t_y\cos k_y$, with $t_x\ne t_y$. To prove this, we consider the
moments
\beq
M^\sigma_n(k)\equiv\int \frac{d\omega}{2\pi} \omega^n d\omega G^{\rm ret}_\sigma(k,\omega)
\eeq
of the Green function.  For simplicity, we have set $\hbar=1$. Using the Heisenberg equations of motion,
we reduce\cite{moments} the moments in real space
\beq
M^\sigma_n(i,j)&=&\frac12\left[\langle\{[H,[H\cdots[H,c_{i\sigma}]\cdots]_{\rm n\, times},c_{j\sigma}^\dagger\}\rangle\right.\nonumber\\
&+&\left.\langle\{c_{i\sigma},[\cdots[c^\dagger_{j\sigma},H]\cdots H],H]_{\rm
  n\, times}\}\rangle\right]
\eeq
to a string of commutators of the electron creation or annihilation
operators with the Hubbard Hamiltonian.  The right-hand side of this expression is evaluated at equal times.  To evaluate the string of commutators, it suffices to focus on the properties of
$K_{i\sigma}^{(n)}=[\cdots[c_{i\sigma},H],\cdots H]_{\rm n \, times},$
where by construction, $K^{(0)}_{i\sigma}=c_{i\sigma}$.  We write the Hubbard Hamiltonian as $H=H_t+H_U$ where $H_U$ includes the interaction as well as the chemical potential terms and $H_t$ the hopping term. The form of the first commutator,
\beq
K_{i\sigma}^{(1)}=\sum_jt_{ij}c_{j\sigma}+Uc_{i\sigma}n_{i-\sigma}-\mu c_{i\sigma}
\eeq
suggests that we seek a solution of the form
\beq
K_{i\sigma}^{(n)}=\sum_j t_{ij}\Lambda_{j\sigma}^{(n)} +Q_{i\sigma}^{(n)}
\eeq
where $Q_{i\sigma}^{(n)}=[\cdots[c_{i\sigma},H_U],\cdots H_U]_{\rm n \, times}$ involves a string containing $H_U$ n times and in
$\Lambda_{j\sigma}$, {\bf $H_t$ appears at least once}.  Our proof hinges on the form of $Q_{i\sigma}^{(n)}$ which we write in general as
$Q_{i\sigma}^{(n)}=\alpha_n c_{i\sigma}n_{i-\sigma}+\beta_n c_{i\sigma}$.
The solution for the coefficients
\beq
\begin{array}{ll} \alpha_{n+1}=(U-\mu)\alpha_n+U(-\mu)^n\\ \beta_n=(-\mu)^n
\end{array}
\eeq
is determined from the recursion relationship $Q_{i\sigma}^{(n+1)}=[Q_{i\sigma}^{(n)},H_U]$.  In the moments, the quantity which appears is
\beq
\langle\{Q_{i\sigma}^{(n)},c^\dagger_{j\sigma}\}\rangle=\delta_{ij}\left[\alpha_n \langle n_{i-\sigma}\rangle+\beta_n\right]\equiv \delta_{ij}\gamma_n.
\eeq
Consequently, the moments simplify to
\beq\label{finalmoment}
 M^{\sigma}_n(i,j)=\delta_{ij}\gamma_n+\frac12\sum_lt_{il}\left(\langle\{
 \Lambda^{(n)}_{l\sigma},c^\dagger_{j\sigma}\}\rangle+h.c.\right).
\eeq

The criterion for the zeros of the Green function
now reduces to a condition on the parity of the right-hand side of
Eq. (\ref{finalmoment}).  Consider the case of half-filling, particle-hole symmetry and nearest-neighbour hopping. Under these conditions, $\langle n_{i\sigma}\rangle=1/2$ and by particle-hole symmetry, $\mu=U/2$.  The expressions for $\alpha_n$ and $\beta_n$
lays plain that the resultant coefficients
\beq\label{gamma}
\gamma_n=\left(\frac{U}{2}\right)^n\frac{1+(-1)^n}{2}
\eeq
vanish for $n$ odd.  Consequently, $G_\sigma(k,\omega)$ is an even function if the second term in Eq. (\ref{finalmoment}) vanishes.  In Fourier space, the second term is proportional to the non-interacting band structure $t(k)$. The momenta at which $t(k)=0$ define the Fermi surface of the non-interacting system.
Note, the condition $t(k)=0$ which defines the surface of zeros {\bf is independent} of the anisotropy of the hopping.  We conclude that when particle-hole symmetry is present, $G(\vec p=\vec p_F,0)=0$ for a Mott insulator, where $\vec p_F$ is the Fermi surface for the non-interacting
 system.  In this case, the volume of the surface of zeros is identically equal to the particle density.   This constitutes
one of the few exact results for Mott insulators that is independent
of spatial dimension or at least as long as $d\ne\infty$.   As mentioned previously, in $d=\infty$, there is no Luttinger surface as $\Sigma$ diverges as $1/\omega$ for all momenta\cite{dmft}.
 Finally, the only
condition for the applicability of our proof is that the form of the
spectral function leads to the continuity of $R_\sigma(\vec
k,\omega)$ at $\omega=0$. Hence, the minimal condition is that the
spectral function is be continuous at $\omega=0$. Therefore, if
there is a gapless quasiparticle excitation, for example, $A_\sigma (\vec k, \epsilon) =
\delta(\omega)$, our proof becomes invalid.

\subsection{Away From Particle-Hole Symmetry}
What happens when particle-hole symmetry is broken?  To consider this regime,
we write the electron density
\beq
n=-2i\sum_k\lim_{t\rightarrow 0^+}\int \frac{d\omega}{2\pi} G(k,i\omega) e^{i\omega t}
\eeq
as an integral of the time-ordered Green function where the factor of two counts up and down spin electrons.  In proving Luttinger's theorem, one uses the identity
\beq
G(k,i\omega) = \frac{\partial}{\partial i\omega}\log G^{-1}(k,i\omega) + G(k,i\omega) \frac{\partial}{\partial i\omega}
\Sigma(k,i\omega).
\eeq
Implicit in this expression is the Dyson equation,
\beq
G^{-1}(k,i\omega)=G_0^{-1}(k,i\omega)+\Sigma(k,i\omega),
\eeq
where $\Sigma$ is the self energy and $G_0$ the Green function for the non-interacting problem.
The density, $n=I_1-I_2$, is now a sum of two terms
\beq\label{eq:I}
I_1 \equiv -2i \sum_k \int_{-\infty}^{\infty}\frac{d\omega}{2\pi} \frac{\partial}{\partial i\omega}\log G^{-1}(k,i\omega)\\
I_2 \equiv 2i \sum_k \int_{-\infty}^{\infty} \frac{d\omega}{2\pi}G(k,i\omega)\frac{\partial}{\partial
i\omega}\Sigma(k,i\omega).
\eeq
Luttinger\cite{luttinger} proved that $I_2$ vanished and hence the electron density is given
simply by Eq. (\ref{density}).  In fact, Dzyaloshinskii\cite{dzy} has claimed that $I_2$ vanishes for a Mott insulator.  Central to this proof is the existence of a perturbative expansion for the self energy around the atomic limit.
Based on the self-energy, the
Luttinger-Ward (LW) functional\cite{luttinger},
\beq \label{eq:LW} \delta \Phi [G] = \sum_k \int \frac{d\omega}{2\pi} \Sigma (k,i\omega) \delta G(k,i\omega)
\eeq
can be constructed which for a Fermi liquid has a perturbative expansion in terms of skeleton diagrams. In general, the LW functional is assumed to have a perturbative expansion.  As such any perturbative LW functional must be free of singularities and vanish as $\omega\rightarrow\infty$.  We show here that for the Mott problem, singularities arise and it is precisely from the singularities that a breakdown of Luttinger's theorem arises.

To see how Luttinger's theorem fails for a Mott insulator,
consider the exact temperature-dependent Green function,
\beq\label{eq:G-t}
G(k,i\omega) &=& \frac{1}{i\omega + \mu + U/2 - \Sigma_{loc}(i\omega)} \nonumber\\
&=&\frac{i\omega+\mu}{(i\omega- E_1)(i\omega - E_2)},
\eeq
in the atomic limit
where
\beq\label{eq:self}
\Sigma_{\rm loc}(i\omega)=\frac{U}{2} + \left( {\frac{U}{2}} \right)^2\frac{1}{i\omega+\mu}
\eeq
and  $E_{1,2} = - \mu \pm U/2$.   As the Mott gap is well-formed in this limit, any conclusion we reach regarding $I_2$ will hold as long as $U\gg t$. Can Eq. (\ref{eq:self}) be constructed from the non-interacting limit?   To all orders in perturbation theory at $T=0$, the self-energy is given by
 \beq
\Sigma_{\rm pert}=U.
\eeq
Such a self-energy cannot describe the two-peak structure of the Mott insulating state.  That is, starting from the non-interacting system, one cannot obtain the Mott gap perturbatively.
It is this breakdown of perturbation theory in generating the Mott gap
that is central to the ultimate breakdown of the Luttinger sum-rule on the volume of the surface of zeros. Given that $\Sigma_{\rm pert}\ne\Sigma_{\rm loc}$,
the corresponding LW functional, $\Phi(G)$ cannot be obtained perturbatively.
Consequently, we must resort to a non-perturbative method to construct the LW functional.  To gain some insight into what the corresponding LW functional looks like, we rewrite $\Sigma_{\rm loc}$ in terms of the exact $G$,
\beq\label{eq:self-2}
\Sigma[G]=\frac{U}{2} + \frac{-1 \pm \sqrt{1+U^2 G^2}}{2G}
\eeq 
by eliminating the $1/(i\omega_n+\mu)$ factor by using the Dyson equation and solving the subsequent quadratic equation.
 Here, the upper and lower signs
should be used when $ |\omega + \mu| > U/2$ and $|\omega + \mu| < U/2$ respectively.  As $\Sigma[G]$ satisfies the Dyson equation, it is exactly given by the functional derivative of the exact LW functional with respect to $G$. Hence, Eq. (\ref{eq:self-2}) implies that we know $\delta\Phi[G]/\delta G$ at the saddle point of $\Phi$. Constructing $\Phi$ in general, however, requires complete knowledge of $\delta\Phi[G]/\delta G$ not simply at the saddle point. How then do we construct $\Phi$? 

For the problem at hand, there are two requirements that any approximate expression for $\Phi[G]$ must satisfy: 1) it must contain a singular part and 2) $I_2$ computed from any approximate
LW functional must agree with a direct calculation based on the second of Eqs. (\ref{eq:I}).  The approximate LW-functional we derive here satisfies both of these requirements and hence lends credence to the method. Faced with the similar problem of obtaining $\Phi[G]$ knowing only its exact derivative at one particular value of $G$ for a spin-density wave problem, Altshuler, et al.\cite{altshuler} simply integrated $\Sigma[G]$ with respect to $G$ to obtain an approximate LW functional. They showed that this procedure to be internally consistent for their problem as $I_2$ evaluated with the approximate LW functional agreed with a direct calculation of $I_2$ from Eq. (\ref{eq:I}). We adopt this approach and check its internal consistency in a similar manner.  The integral
of $\Sigma[G]$, 
\beq\label{eq:phi}
\Phi(i\omega)&=& \frac{1}{2} \left[ -\log G(i\omega) \pm \sqrt{1+U^2 G^2(i\omega)}\right. \nonumber\\
&&\left.\pm \frac{1}{2} \log \left(  \frac{\sqrt{1+U^2 G^2(i\omega)}-1}{\sqrt{1+U^2 G^2(i\omega)}+1}  \right) \right]\nonumber
\\&=& \Phi_{\rm reg}(i\omega) + \Phi_{sing}(i\omega),
\eeq
contains both a regular as well as a singular part,
\beq\label{eq:LW-sing}
\Phi_{\rm sing}(i\omega)= \frac{1}{2} \log \frac{G_0 (i\omega)}{G (i\omega)}.
\eeq
Although the atomic limit differs from the spin-density wave problem treated by
Altshuler, et al.\cite{altshuler}, the approximate LW functionals are identical. This state of affairs obtains because of the similarity between the self energies
of the two problems.
To evaluate $I_2$, we note that only the singular part of $\Phi[G]$ contributes. The result
\beq\label{eq:I2}
I_2&=&-2i\int\frac{d\omega}{2\pi} \frac{\partial \Phi_{sing}(i\omega)}{\partial i\omega}\nonumber\\
&=&2\Theta(\mu) - \Theta(-E_1) - \Theta(-E_2),
\eeq
 is in agreement with a direct calculation\cite{rosch} of $I_2$ based on the second equation in Eq. (\ref{eq:I}).  This agreement suggests that our approximate expression for $\Phi[G]$ captures the essence of the breakdown of Luttinger's theorem. Note $I_2$ term vanishes only in the presence of particle-hole symmetry ($\mu=0$). The modified Luttinger theorem becomes
\beq\label{eq:Lutt-general}
n = \int_{G(0,k)>0}\frac{d^2k}{(2\pi)^2} + \Theta(\mu).
\eeq
The modified Luttinger theorem is also valid even in the presence of small hopping, that is, $U\gg t$,
as can be seen by considering
\beq\label{eq:G-tt}
G(k,i\omega) &=& \frac{1}{i\omega - t (k) + \mu  + U/2 - \Sigma_{loc}(i\omega)} \nonumber\\
&=&\frac{i\omega+\mu}{(i\omega- E_1(k))(i\omega - E_2(k))} \eeq
for the Green function. Here $t(k)$ is the Fourier transformation of the hopping element, $t_{ij}$. Substitution of $G$ and $\Sigma$ into the second term of
Eq. (\ref{eq:G-tt}) leads to
\beq\label{eq:I22} I_2= 2\Theta(\mu) -\int \frac{d^2k}{(2\pi)^2} [\Theta(-E_1(k))
+ \Theta(-E_2(k))].
\eeq
Once again, $I_2$ vanishes in the presence of particle-hole symmetry and hence can be rewritten as Eq. (\ref{eq:I2}). 

The form of the singular part of the Luttinger-Ward functional compels a simpler formulation of the electron density.
Since the regular part of $I_2$ vanishes, we can use Eqs. (\ref{eq:I}) and (\ref{eq:LW-sing}) to recast the electron density as a sum of two contributions,
\beq\label{atomlim}
n&=&i\sum_k \int_{-\infty}^{\infty}\frac{d\omega}{2\pi} \frac{\partial}{\partial i\omega}\log G(k,i\omega)G_0(k,i\omega)\,\,\,\,\nonumber\\
&\equiv&\int_{G(k,\omega=0)>0}\frac{d^2k}{(2\pi)^2}+\int_{G_{t=0,U=0}(k,
\omega=0)>0}\frac{d^2k}{(2\pi)^2},\nonumber\\
\eeq
each of the Luttinger form.  Although both of the terms in this expression (as well as in Eq. (30)) contain discontiuities, the discontinuities cancel in the sum leading to the density being a continuous function of the chemical potential.  Eq. (\ref{atomlim}) is valid  in the atomic limit  as well as in the small hopping regime and represents the general form of Luttinger's theorem for a Mott insulator. Note in the weak hopping limit, only the first term differs from that in the atomic limit. In interpreting Eq. (\ref{atomlim}), it is important to remember that the second term is not equivalent to $I_2$.  Part of $I_2$ cancels one of the Green functions in the first term.  The term which is left over accounts for the fact that the chemical potential can be placed arbitrarily within
the gap as emphasized previously\cite{rosch}.  This ambiguity, of course, is absent for a soft gap as in the case of the pseudogap in the doped case. In this case, however, the exact self-energy is not known and no recasting of the Luttinger theorem as the general statement in Eq. (\ref{atomlim}) is possible.  What the current analysis lays plain is that the singular part of the LW functional, which is absent for a Fermi liquid, leads to the break-down of the Luttinger sum rule on the surface of zeros in the absence of particle-hole symmetry. 

The current analysis can be extended to finite temperature.  At finite $T$, $I_2$,
\beq
\label{finiteT}
I_2(T>0)&=&f(-\mu+\frac{\sqrt{U^2+\varepsilon^2(k)}}{2})\nonumber\\
&&+f(-\mu-\frac{\sqrt{U^2+\varepsilon^2(k)}}{2})-2f(\mu),
\eeq
can be evaluated using Matsubara
frequencies and the singularity which originally existed on the real
frequency axis can be removed.  Eq. (\ref{finiteT}) goes over smoothly to the
zero-temperature $I_2(T=0)$ evaluated by
real frequency integration. Therefore, the non-vanishing of $I_2(T=0)$
is independent of the regularization of the
singular part of the LW functional and is a generic feature of a
Mott insulator.
  This result is significant because Dzyaloshinskii\cite{dzy} proposed that the non-vanishing of $I_2$ stemmed from the method Altshuler, et. al.\cite{altshuler} used to regularise the singular integrals.  Namely that the breakdown of Luttinger's theorem for a spin-density wave arises from a $T=0$ regularisation of divergent integrals that can only be reached if there is a $T=0$ phase transition.  The current work establishes that there is no such phase transition and the $T=0$ result is connected adiabatically to the finite-T result.  Consequently, the breakdown of Luttinger's theorem lies elsewhere.  In both the Mott insulator and spin-density wave problems, no perturbative expansion exists for the self-energy around the non-interacting limit. Should this fail, there will always be a singular part of the LW functional and $I_2$ will be finite.  Altshuler et. al.\cite{altshuler} made a connection between such a break down and the chiral anomaly in particle physics. While at the atomic limit, this association might be appropriate, it is unclear whether this analogy holds for the general case.   To reiterate, for a Mott system lacking particle-hole symmetry
but possessing a divergent self-energy, the singular part of $G\partial_\omega\Sigma$
will always integrate to a non-zero value.

\subsection{Consequences}
Several claims follow necessarily from these results.\\
C1.\quad {\it There are no non-trivial zeros of the single-particle Green function in the {\bf single-impurity} Anderson model.}
Because the gap is replaced by the Kondo resonance, no zeros of the Green function obtain for the single-impurity problem.\\
C2.\quad {\it At the surface of zeros, the self-energy at zero frequency diverges.}
Write the single-particle time-ordered Green function as
$G_\sigma(k,\omega)=1/(\omega-\epsilon(k)-\Re\Sigma_\sigma(k,\omega)-i\Im\Sigma_\sigma(k,\omega))$,
where $\Sigma$ is the self-energy.
 For a particle-hole symmetric band structure, the single-particle Green function vanishes
linearly at the Luttinger surface, $k_L$: $G_\sigma(k,\omega=0)={\rm const.}\times(k-k_L)$.
This implies that
\beq\label{div1}
\Im\Sigma(k,0)\propto\delta(k-k_L).
\eeq
Note, however,
that $\Im G(k,\omega)=0$ for all energies within the gap. By inverting the Green
function, it follows that
\beq\label{div2}
\Re\Sigma_\sigma(k,\omega)\propto (k-k_L)^{-1}
\eeq
 proving C2.  To reiterate, in $d=\infty$\cite{dmft} no Luttinger surface exists as the self-energy diverges for all momenta at $\omega=0$.  While such a divergence is appropriate for $d=\infty$, it is clearly unphysical for a finite-dimensional system.  The divergence in Eq. (\ref{div2}) prevents the renormalized energy band $E(k)=\epsilon(k)+\Re\Sigma_\sigma(k,\omega)$ from crossing the Fermi energy. The result is an insulating state.  Indeed, in numerical studies\cite{jarrel2} on the Hubbard model at half-filling with nearest-neighbour hopping, $\Sigma_\sigma(\pi,0,\omega=0)$ has been observed to diverge as our theorem indicates it must. However,
Jarrell and co-workers\cite{jarrel2} attributed antiferromagnetism as the cause of the divergence.  Our theorem indicates that the zeros are independent of the ground state (be it ordered or not as in the case of a spin liquid) as long as the Mott gap is present. The zeros are a direct consequence of Mottness itself.   That Mottness and zeros are one and the same indicates that the divergence of the self energy provides a general mechanism for insulating states in the absence of broken symmetry.  That is, in a finite-dimensional lattice, the divergence of the self-energy at a contiuously connected momentum surface is the general mechanism by which the Mott insulating state obtains.\\
C3.\quad{\it  Zeros represent a breakdown of weak-coupling
perturbation theory.}
This follows directly from C2. A divergence of the self-energy is the general
signature of the breakdown of perturbation theory.  Zeros offer a concrete way of realising this breakdown. In $d=1$, this breakdown
occurs for all $U\ne 0$. In $d=2$ in the particle-hole symmetric case,
the critical value of $U$ is not known, though all numerics\cite{MouJar,stanescu1} indicates that the only special point is $U=0$.\\
C4.\quad {\it The surface of zeros of the single-particle Green function is absent from projected models at half-filling.}
Since it is common in the study of doped Mott insulators to use projected models, it instructive to evaluate whether such truncations admit a surface of zeros.  Projecting out double occupancy, as in the $t-J$ model,
erases the spectral weight above the chemical potential at half-filling.
 Consequently, the real part of the
Green function reduces to the first integral in Eq. (\ref{real}), which is always non-zero. C4 is thus proven.  Transforming the operators in the $t-J$ model to respect the no double occupancy condition is of no help as the problem stems from
the loss of spectral weight above the gap once projection occurs. As the surface of zeros occurs at zero energy and is located in momentum space, it should certainly be present in a low-energy theory of the Hubbard model. However, it is clear from Eq. (\ref{real}) that zeros of the Green function stem from a sum rule\cite{noncomm} connecting low and high energies. Hence, it is {\it a priori} expected that the zero surface would be sensitive to the retention of the spectral weight at high energies.

The absence of zeros in the $t-J$ model at half-filling is in actuality related
 to the problem of the robustness of zeros and the location of the chemical potential at half-filling in a Mott insulator.  At $T=0$, the chemical
potential is a free parameter that can be located anywhere in the Mott gap.  Consider, the extreme case of placing the chemical potential atop the lower Hubbard band and sending $U$ to infinity. In this case, the integrand in the second term in Eq. (4) has an infinite energy denominator and hence the second term vanishes.  Consequently, there are no zeros in this case.  The actual realisation of this is the $t-J$ model at ``half-filling.''\cite{halffilling}  Hence, there are certain locations for the chemical potential for which the zero line vanishes.  This does not diminish the significance of the zero line as the defining feature of a Mott insulator, however.  That the chemical potential is arbitrary at $T=0$ indicates that the $T=0$ value of the chemical potential is not a defining feature of a Mott insulator.
 What is the defining feature of a Mott insulator is that at half-filling, half the spectral weight lies above the gap.  Such a schism in the spectral weight guarantees that the real part of the Green function must change sign along some momentum surface for some energy or range of energies within the gap as emphasized by Dzyaloshinski\cite{dzy}. Our claim that the surface of zeros defines the Mott insulator is simply that dynamical generation of a gap, which at half filling results in half the spectral weight lying above and below the gap, leads to a sign change of $\Re G$ for some (not necessarily all) energies within the gap.  In this vein, the $t-$J model is not a realistic model of the Mott state because a zero line is strictly absent.\\
C5.\quad {\it Even at infinitesimal doping, the t-J and Hubbard models probably do not yield equivalent values for the quasiparticle weight.}  Because the chemical potential sits atop the lower Hubbard band in the $t-J$ model at ``half-filling''\cite{halffilling}, perhaps the proper way to compare
the with the Hubbard model is in the limit of infinitesimal doping (see Figs. (\ref{options}a) and (c)). Numerical and analytical studies on the one hole system\cite{z1,z2,z3} find
 a quasiparticle in the t-J model with weight $J/t$ at $(\pi/2,\pi/2)$ whereas in the Hubbard model\cite{sorella}, the quasiparticle weight vanishes as $Z\propto L^{-\theta}$, $\theta>0$, $L$ the system size. Variational calculations\cite{variational} also yield a finite value of $Z$ in the extrapolated limit of $n=1^-$.  While none of this constitutes a proof, it is highly suggestive that the value of $Z$ is tied to the presence of the upper Hubbard band as has been emphasized previously\cite{noncomm,anderson}. In the one-hole system, sufficient spectral weight must lie above the chemical potential for Eq. (4) to vanish. There is no guarantee that this state of affairs obtains for the $t-J$ model since no spectral weight was above the gap at the outset. No such problem arises for the Hubbard model.  \\
C6.\quad{\it If a Mott gap opens, zeros of the single-particle Green function still persist when the particle-hole symmetry is broken weakly.} As remarked earlier, all that is necessary to establish is that for some energies within the gap, the real part of the Green function changes sign.  At present,
our proof applies to any kind of band structure that is generated from
hopping processes which remain unchanged  after the application of Eq. (\ref{ph}).  In general, the two kinds of hopping processes
transform as $\epsilon(\pi-k_x,\pi-k_y)=-\epsilon(k_x,k_y)$ and $t'(\pi-k_x,\pi-k_y)=t'(k_x,k_y)$. The latter describes next-nearest neighbour hopping and as is present in the cuprates.  If only such hopping is present, the
surface of zeros is no longer the diagonal $(\pi,0)$ to $(0,\pi)$ (or the point
$\pi/2$ in 1D),
but rather the ``cross'' $(0,\pi/2)$ to $(\pi,\pi/2)$ and $(\pi/2, 0)$ to $(\pi/2,\pi)$ (or, in 1D, the
points $-\pi/4$ and $3\pi/4$).  When both types of hopping are present, no symmetry arguments can be made.  Our proof in this case will rely on a key assumption: the Green
function is a continuous function of the hopping parameters $t$ and $t'$. Hence, strictly speaking our proof applies only when $t`\ll t$. When only
$t$ is present, $R_\sigma(k,0)$ has one sign (plus) near $k=(0,0)$
(or, in 1D, k=0), and the opposite (minus) near
$k=(\pi,\pi)$ ($k=\pi$ in 1D)
and will vanish on the zero line. Alternatively, if we have $t'$ hopping,
$R_\sigma(k,0)$
will have a certain sign near $k=(0,0)$ and $k=(\pi,\pi)$,
and the opposite sign
near $k=(0,\pi)$ and $k=(\pi,0)$ and will vanish on the "cross".
From continuity, for $t'\ll t$, $R_\sigma(k; t, t')$ will have the same
sign structure as $R_\sigma(k;
t, t'=0)$. That is, it will change sign when going from $(0,0)$ to $(\pi, \pi)$
regardless of the path taken.  Therefore, the line of zeros exists
for small enough t', the relevant limit for the cuprates. In the opposite limit, $t'\gg t$, a similar argument holds. Whether a proof exists for the case of a
strong violation of particle-hole symmetry is not known.

\section{Utility of Zeros: Pseudogap phase of  Doped Mott Insulators}
Ultimately, the utility of the surface of zeros will be determined by its experimental relevance.  As we have seen above, the volume of the surface of zeros can only be calculated explicitly in the case of particle-hole symmetry. At finite doping where this symmetry is explicitly broken, the existence (though not the volume) of the zero surface can nonetheless be established.  Two independent arguments are relevant here. The first is based on the distribution of spectral weight in the spectral function and the other on the fact that
Fermi arcs, as seen experimentally\cite{norman,kanigel} in the doped cuprates, necessitate the existence of a surface of zeros.  The general arguments made here for the interdependence of zeros and Fermi arcs augment the numerical evidence found by Stanescu and Kotliar\cite{stanescu4} for the same effect.

\begin{figure}
\centering
\includegraphics[width=7.0cm,angle=-90]{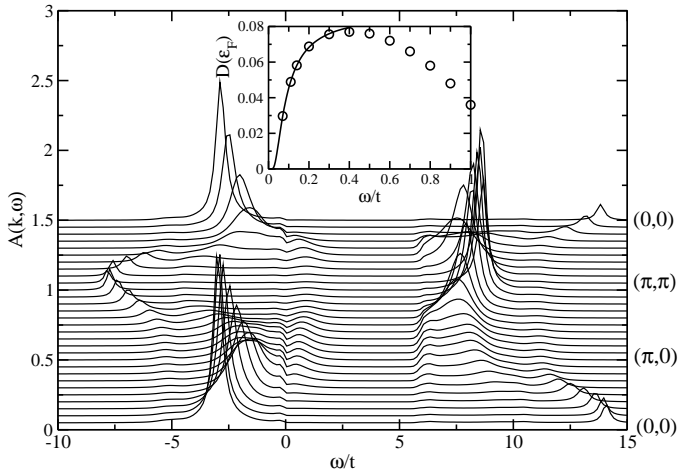}
\caption{Spectral function for a doped Mott insulator at a filling of $n=0.95$ at $T=0.07t$ for a path in momentum space from $(0,0)$ to $(\pi,0)$ to $(\pi,\pi)$ and then back to $(0,0)$. Eq. (\ref{real}) must undergo
a sign change by passing through zero because at $(0,0)$ most of the spectral weight lies below the chemical potential, whereas at $(\pi,\pi)$, it lies above. The temperature dependence of the density of states at the Fermi level is shown in the inset. That the spectral features are broad near the chemical potential is a direct consequence of the divergence of the self energy at $\omega=0$ at the Luttinger surface. The spectral function was computed using the two-site self-consistent method of Stanescu and Phillips\cite{stanescu1}.``Color Online''}
\label{zeros}
\end{figure}

To illustrate that the spectral weight distribution in a lightly doped Mott insulator supports a zero surface, we consider the spectral function shown in Fig. (\ref{zeros}). The computational scheme used to produce this spectral function has been detailed elsewhere\cite{stanescu1} and is in agreement with results from state-of-the-art\cite{jarrel2} calculations on the 2D Hubbard model.  Two features are relevant.  First, it possesses a depressed density of states at the chemical potential for a wide range of momenta.  This leads to a density of states which vanishes algebraically at the chemical potential, as is seen experimentally\cite{asym}.  Such a dynamically generated pseudogap which occurs without any symmetry breaking has been confirmed by all recent numerical computations on the the doped Hubbard model\cite{stanescu3,c2,c3,stanescu4,jarrel2}. Hence, that $\Im G(0,\vec p)=0$ along some contour in momentum space for a doped Mott insulator is not in dispute neither theoretically nor experimentally. What about $\Re G(0,\vec p)$?  As is clear from Eq. (\ref{real}), $\Re G(0,\vec p)=0$ if along some contour in momentum space, the spectral weight changes from being predominantly below the chemical potential to lying above.   At half-filling, the zero surface obtains
entirely because most of the spectral weight at $(\pi,\pi)$ lies
above the chemical potential, whereas at $(0,0)$, it lies below.
As is evident, this trend still persists for $x\approx 1$ as Fig. (\ref{zeros})
attests.  Hence, $R_\sigma(k,0)$ still has the same sign structure
as in the undoped case.  Consequently, a zero surface must exist. 

\begin{figure}
\centering
\includegraphics[width=7.0cm,angle=0]{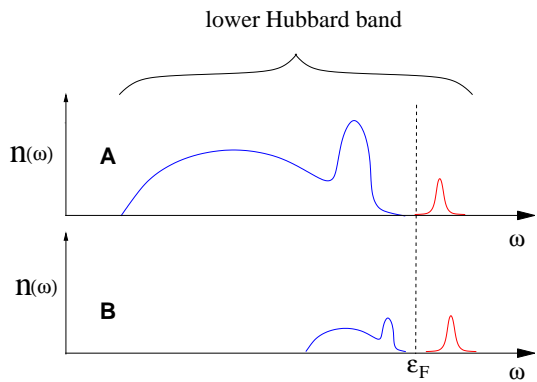}
\caption{Schematic depction of two possibilities for the distribution
of spectral weight in the single-particle density of states, $n(\omega)$, in a lightly doped Mott insulator. In both cases, the peak above the chemical potential represents the low-energy spectral weight.  The weight of this peak increases as least as fast as $2x$, where $x$ is the number of holes. A) The Luttinger surface lies along the zone diagonal and the transferred spectral weight must be asymetrically located relative to the chemical potential to lead to a vanishing of $\Re G$. B) The spectral weight redistribution is symmetrical and the Luttinger surface lies away from the zone diagonal, for example close to $(\pi,\pi)$. }
\label{specopt}
\end{figure}

Ultimately, satisfying the zero condition, Eq. (\ref{real}),
requires spectral weight to lie immediately above the chemical
potential.  Spectral weight transfer\cite{swt} across the Mott gap
is the mediator.  The weight of the peak above the chemical
potential scales as $2x+f(x,t/U)$\cite{swt} (strictly 2x in the t-J
model) while the weight below the chemical potential is determined
by the filling, $1-x$.  Whether or not the redistributed spectral weight is symmetric or not around the chemical potential will determine how severely the Lutinger volume is violated.  There are only two options as depicted in Fig. (\ref{specopt}). \\
{\bf Weak violation of Luttinger volume:} In order to satisfy Luttinger theorem, the surface of zeros has to be close
to the $(0,\pi)-(\pi,0)$ line (assuming that $t$' is small). Consider a point on the zero line in the vicinity of $(0,\pi)$.  As is evident from Fig. (\ref{zeros}) the spectral weight immediately
above $\mu$ (in the vicinity of the $(\pi,0)-(0,\pi)$ line is small compared with the spectral weight below $\mu$ and, in
order for $\Re G$ to vanish, the chemical potential has to be positioned
asymmetrically inside the pseudogap (see panel A of Fig. (\ref{specopt}).\\
{\bf Strong violation of Luttinger volume:} The other possibility is that the Luttinger theorem is strongly violated and the surface of zeros is somewhere in the vicinity of $(\pi, \pi)$. In that
region, the spectral weight of the lower Hubbard band is greatly reduced as shown in Fig. (\ref{zeros}).
Consequently,  most of the spectral weight lies in the upper Hubbard band
 and a more symmetrical distribution around the chemical potential is possible (see panel B in
Fig. (\ref{specopt}).  

To decide between options A and B in Fig. (\ref{specopt}) we appeal to experiments. Two observations support option A. First, the pseudogap is in fact asymmetrical\cite{asym}.  Second, consider the recent photoemisison experiments\cite{kanigel} in which the temperature dependence of the Fermi arcs has been measured.  Experimentally, lightly doped cuprates possess Fermi arcs\cite{norman} along the zone diagonal in the vicinity
of $(\pi/2,\pi/2)$.  Whether the Fermi arcs represent a finite $T$ precursor of a Fermi surface and hence quasiparticles as claimed by some\cite{rice} can be settled by temperature-dependent ARPES experiments in the pseudogap regime.
 Kanigel\cite{kanigel} et al. performed such temperature-dependent ARPES measurements on Bi$_2$Sr$_2$CaCu$_2$O$_{8+\delta}$ (Bi2212) and concluded that the Fermi arc length shrinks to zero as $T/T^\ast$, where $T^\ast$ is the temperature at which the pseudogap feature appears. Hence, the only remnant of the arc at $T=0$, is a quasiparticle in the vicinity of $(\pi/2,\pi/2)$. Consequently, Kanigel, et al.\cite{kanigel} argue for a nodal metal.  That a nodal metal or a quasiparticle band existing over a finite connected region in momentum space not extending to the zone boundary (as would be the case in a $T=0$ Fermi arc) cannot be understood without the existence of a surface of zeros can be seen as follows. Assume a quasiparticle exists at $(\pi/2,\pi/2)$.  Then $\Re G(0,p)$ must change sign for all momenta less than or greater than $(\pi/2,\pi/2)$ as depicted in Fig. (\ref{options}C). Consider traversing a path through  $(\pi/2,\pi/2)$ and then returning along a path that does not cross this point.  To end up with the correct sign for $\Re G$, the return path must intersect a line across which $\Re G(0,\vec p)$ changes sign. Since there are no infinities, except at $(\pi/2,\pi/2)$, the only option is for a zero line to exist. The zero line must emanate from the $(\pi/2,\pi/2)$ point and touch the edges of the Brillouin zone close to $(\pi,0)$ and $(0,\pi)$. A zero surface terminating close to $(\pi,\pi)$ is not an option as this would permit the existence of paths that traverse the zone diagonal without changing the sign of $\Re G(0,\vec p)$.
This would suggest that the zero surface in the doped cuprates preserves the Lutinger volume and option A in Fig. (\ref{specopt}) is more consistent with experiment.  As a consequence, Fermi arcs are direct evidence that zeros of the single particle Green function must be present in the doped cuprates.  In a recent paper, Stanescu and Kotliar\cite{stanescu4} have argued based on numerics for such an interdependence.
\section{Concluding Remarks}
\begin{figure}
\centering
\includegraphics[width=8.5cm]{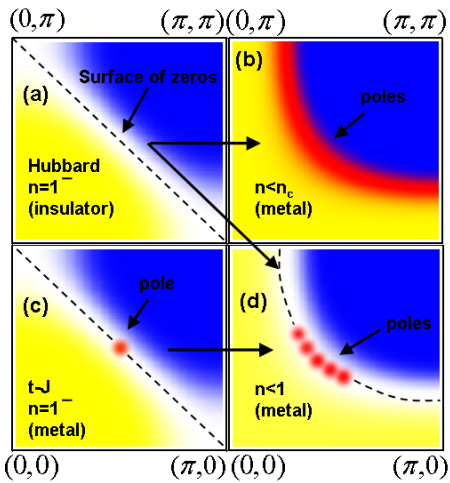}
\caption{(Color online) Evolution of the surface of zeros in the first quadrant of the FBZ. Yellow indicates $\Re G>0$ while blue $\Re G<0$. The Hubbard model is constrained to have a surface of zeros as $n\rightarrow 1$ whereas the t-J model is not. The two options upon doping represent
weak-coupling (1a or c to d) and strong-coupling (1a to b). The transition from (a) to (b) requires a critical point at $n_c$ whereas (d) does not. Experiments\cite{boeb,ando} indicating an insulating
state for $n>n_c$ are consistent with an abrupt transition from (a) to (b).}
\label{options}
\end{figure}

As we have seen the experimental utility of zeros of the single-particle Green function is in their relevance to ARPES.  One of the hallmarks of the normal state of the cuprates is an absence\cite{norman} of electron-like quasiparticles.  Quasiparticles require a vanishing of the renormalized band, $E(k)=\epsilon(k)+\Re\Sigma (k,\omega=0)$; but because $\Re\Sigma$ diverges along the surface of zeros, no quasiparticles form and broad spectral features
are inevitable as seen in ARPES in the cuprates\cite{norman}.  The clearest experimental signature that the surface of zeros exists is the recent temperature- dependent ARPES experiments that indicate that the Fermi arcs shrink to a point as $T\rightarrow 0$.  Since the surface clearly exists for the cuprates, the
only outstanding question is how does the surface of zeros evolve as a function of doping.  Various options are shown in Fig. (\ref{options}). An abrupt transition from a surface of zeros to a Fermi surface would describe a transition from an insulator to a metal.  Such a transition would require a phase transition at $x_c$, the doping level at which the pseudogap terminates.  Alternatively, quasiparticles and zeros could co-exist.  While we have advocated the former scenario based on a calculation of the conductivity which reveals that the pseudogap is an insulating gap\cite{noncomm2}, a result consistent with experiment\cite{boeb,ando} ultimately, both scenarios are possible, in principle\cite{rice2}.  The former corresponds to an insulator (or nodal metal) whereas the latter describes a metal.  The recent angle-resolved photoemisison experiments\cite{kanigel} seem to indicate that the only co-existence of quasiparticles and zeros occurs at a single point indicating that a Fermi surface is possible only for some doping level exceeding $x_c$ as depicted in the upper panels in Fig. (\ref{options}).

\acknowledgements We thank the NSF, Grant No. DMR-0305864 and G. Kotliar for a helpful e-mail exchange and J. C. Campuzano for a discussion regarding his experimental data. TDS was funded partially by the Bliss Faculty and University Scholar funds from the College of Engineering at UIUC and NCSA.


\begin{thebibliography}{99}
\bibitem{halffilling}Here the quotations indicate that the $t-J$ model at half-filling is in fact ill-defined. In the Mott insulator, the chemical potential lies somewhere in the gap. At infinitesimal doping, it jumps to eithe the top or bottom of the lower and upper Hubbard bands, respectively. In the $t-J$ model, it is located strictly at the top of the lower Hubbard band at ``half-filling.'' Hence, half-filling in the $t-J$ model really corresponds to $n=1^-$, where $n$ is the filling.
\end{thebibliography}
\end{document}